\pdfoutput=1
\RequirePackage{ifpdf}
\documentclass{JINST}
\usepackage{cite}

\title{PMT Triggering and Readout for the MicroBooNE Experiment}

\author{D. Kaleko$^a$, for the MicroBooNE Collaboration
~\\
\llap{$^a$}Columbia University,\\
  538 W 120th St., New York, NY, USA\\
E-mail: \email{kaleko@nevis.columbia.edu}}

\abstract{This paper presents the proposed PMT readout and triggering system that will be used in the MicroBooNE LArTPC experiment. The triggering scheme has been designed to study beam neutrino events as well as fully characterize cosmic rays. In addition, exploration of important physics applications including "late" scintillation light in argon and Michel electrons from muon decay will be possible. Various types of triggers and how they will be implemented in the combined PMT+TPC readout electronics system will be discussed. }


\begin{document}

\section{General MicroBooNE}\label{sec:generalub}

MicroBooNE is a neutrino experiment under construction at Fermi National Accelerator Laboratory (FNAL). It employs a 10m$\times$2.6m$\times$(2.5m drift length) 86 ton active volume (170 ton total volume) liquid argon time projection chamber (TPC) to record ionization signals from particles produced in neutrino interactions, and uses scintillation light detected by a photomultiplier tube (PMT) array to provide precise interaction timing information as well as event-overlapping cosmic ray identification and possible particle identification capabilities. The TPC employs 3mm wire spacing, 2.5m drift (1.6ms drift time), and three wire readout planes. The optical system consists of 32 8-inch cryogenic Hamamatsu PMTs equipped with wavelength-shifting tetraphenyl butadiene (TPB) coated acrylic front plates.\\\\
Located on-surface, the MicroBoone detector will be in the path of the Booster Neutrino Beamline (BNB), a conventional neutrino beam at Fermilab with peak neutrino energy around 800 MeV, as well as the NuMI off-axis beamline with peak neutrino energy around 2 GeV. The BNB beam spill has a duration of 1.6$\mu s$, and a beam spill frequency of 5-15 Hz. The background cosmic rate in the detector is expected to be on the order of 3-5kHz, which corresponds to 5-8 cosmics expected per 1.6ms drift time.

\section{PMT Readout Goals and Limitations}\label{sec:goals}
The PMT readout system has a number of physics goals to accomplish, all while maintaining a neutrino event trigger rate and therefore data rate that the electronics can process and store.

\subsection{PMT Trigger}\label{sec:yyy}
One of the primary purposes of the PMT system in the MicroBooNE experiment is to provide a prompt scintillation light based trigger on events in time coincidence with the BNB or NuMI beam gates. This is useful to significantly reduce the number of events read out for analysis. For example, without a PMT trigger, 1 in 2500 BNB beam spills read out is expected to contain a neutrino event, assuming 2$\times10^{12}$ protons per spill. With a PMT trigger, this ratio is expected to be increased to approximately 1 in 20, thus rejecting empty spill data.

\subsection{Background Cosmic Rays}\label{sec:yyy}
The TPC data readout window when triggered is 4.8 milliseconds, during which time 15-24 background cosmics are expected to enter the detector. It is vital to locate these cosmics in space and time to tag them efficiently during the reconstruction process. The PMT readout structure is designed such that cosmics are read out during the entire TPC readout window duration without surpassing data rate limitations in the readout electronics.

\subsubsection{The ``Flash Finder" Algorithm}
The ``Flash Finder" (FF) is an offline algorithm implemented during the event reconstruction process that uses PMT signals to identify a subevent in time with the beam, to help locate a neutrino interaction amidst multiple cosmic subevents in the same TPC readout window. FF uses pulse timing and amplitudes on multiple PMTs to provide position and timing information for cosmics to be rejected in the reconstruction process. The PMT readout system is designed such that the FF algorithm has all of the input information it needs to isolate cosmic backgrounds and identify neutrino vertex information in the reconstruction process.

\subsection{Prompt and Late Scintillation Light for Particle Identification}\label{sec:yyy}
Scintillation light in liquid argon emitted at 128nm is produced via two pathways, involving molecular excimer states excited either in a singlet state or a triplet state ~\cite{promptlatelightref}. These two pathways have lifetimes that differ by orders of magnitude; the singlet (fast) lifetime is $\tau_F\approx6ns$, while the triplet (slow) lifetime is $\tau_S\approx1.6\mu s$. The quenching of scintillation light is dependent upon local excimer concentrations as a result of a competing excimer collision process, and is therefore dependent on the dE/dx of the ionizing particle ~\cite{benref}. This process more strongly quenches the slow light, which implies that the ratio of prompt to delayed light may be useful in identifying the ionizing particle type. In order to exploit this possibility, the PMT readout system is designed such that readout windows surrounding the beam gate extend sufficiently to view the late light for particle identification purposes.

\subsection{Michel Electron Identification}\label{sec:yyy}
Michel electrons are those created following a stopped muon decay, which has a lifetime of 2.2$\mu s$ ~\cite{pdgref}. Identifying a Michel electron at the end of a track in the detector is useful in identifying the parent particle as a muon. Michel electrons are also useful in PMT quantum efficiency tube-to-tube calibration because of their sharp endpoint energy~\cite{michelpaper}~\cite{behrendspaper}~\cite{kinoshitapaper}, and calibrating light yield as a function of position in the detector. The PMT readout system is designed such that readout windows surrounding the beam gate extend sufficiently long to view scintillation light from a Michel electron following a muon decay inside of the detector.

\section{PMT Electronics Specifics}
\begin{figure}[tbp] 
\centering
\includegraphics[width=.8\textwidth]{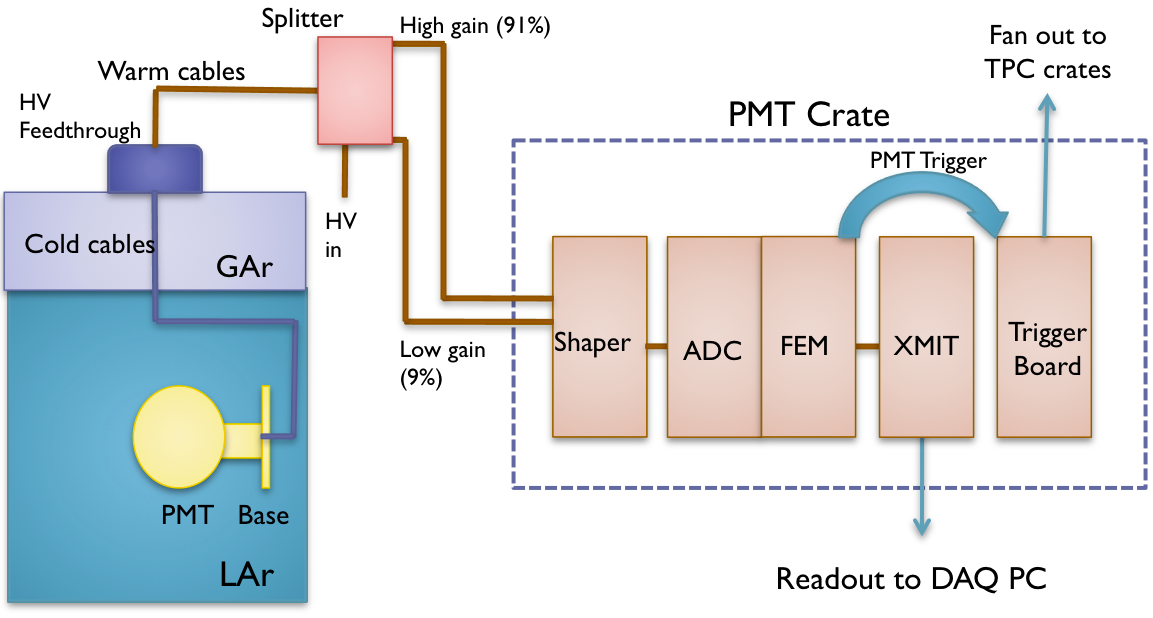}
\caption{The MicroBooNE PMT Readout System.}
\label{fig:electronics_schematic}
\end{figure}
The PMT electronics system is set up as diagrammed in Figure \ref{fig:electronics_schematic}. The splitter boards separate the PMT signals from the HV and provide a high gain and a low gain channel carrying respectively 91\% and 9\% of the signal to the shaper. The purpose of the high/low gain is to extend the dynamic range of the ADC digitizing boards. The analog shaper is unipolar, with 60ns shaped rise time, which allows for accurate event start time determination. Shaped pulses are driven differentially to the 12-bit ADC, where they are digitized at 64MHz (16ns samples). Digitized pulses are fed into a front end module (FEM), on which there is an FPGA responsible for applying PMT readout conditions and generating PMT triggers. PMT triggers are communicated over an external cable to a trigger board, which applies trigger logic and issues a trigger to the PMT crate as well as to all TPC readout crates for event data readout. All data are read out over optical fibers via an XMIT transmitting module.

\subsection{PMT Readout Structure Design}
\begin{figure}[tbp] 
\centering
\includegraphics[width=.8\textwidth]{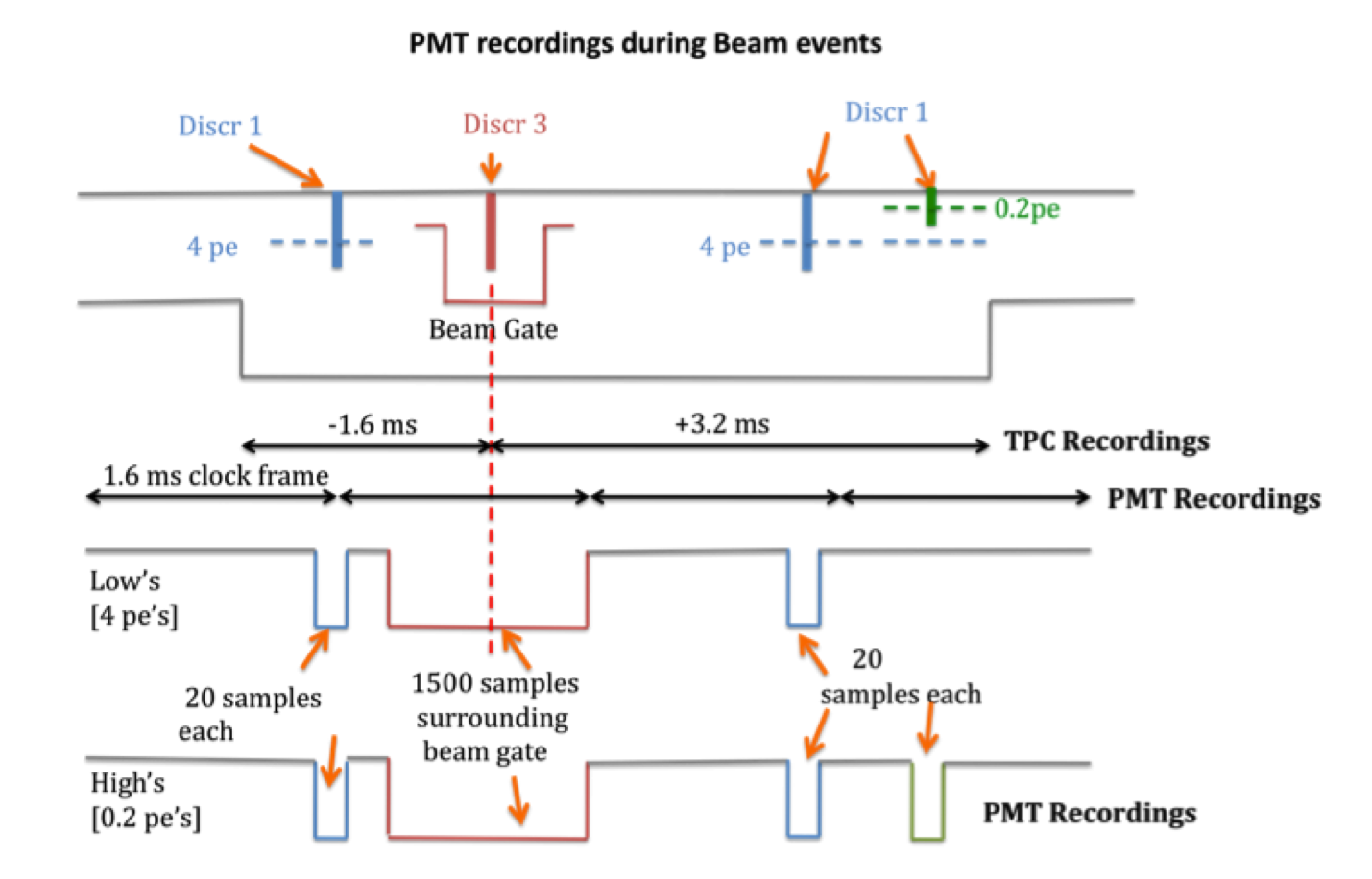}
\caption{Example PMT readout structure for a beam gate trigger event. From top to bottom: signals in the detector with assumed Discr1 threshold of 4 photoelectrons (pe) in the ``low gain" channel and 0.2 pe in the ``high gain", TPCW, ``low gain" channel readout with 1500 sample SW, ``high gain" channel readout with 1500 sample SW. The trigger issued by the trigger board is located inside of the beam gate.}
\label{fig:pmt_readout}
\end{figure}

To achieve the physics goals outlined in section \ref{sec:goals} while maintaining a manageable data rate from the PMTs, various readout windows need to be defined.
\begin{itemize}
\item BGW: Beam Gate Window (1.6$\mu s$ for BNB)
\item SW: Surrouding Window (preceeding BGW start by 4$\mu s$, extending for 24$\mu s$ [1500 digitized samples])
\item TPCW: TPC Readout Window (4.8ms)
\item OUTW: Outer Window (everything outside of the TPCW)
\end{itemize}
The SW is designed to extend several muon lifetimes for the identification of Michel electrons after the end of the BGW, which is also a long enough duration to see late scintillation light from sub-events in the BGW. The SW preceeds the BGW long enough to tag background cosmics arriving shortly before the start of the BGW, that may deposit scintillation light in the BGW and/or Michels that are created inside the BGW. The SW in its entirety is read out for every beam event trigger issued.\\
It is also necessary to record cosmic rays that arrive inside the TPCW, but outside of the SW for background rejection purposes. In this time frame, 20 digitized samples are read out from each cosmic that arrives in the detector. A timing diagram of a possible readout from a beam event trigger is shown in Figure \ref{fig:pmt_readout}.

\subsubsection{PMT Readout Conditions}
The PMT system implements various discriminator timing conditions for readout. A low-threshold ``Discr0" allows for good timing, and is a precondition for higher threshold ``Discr1" and ``Discr3" discriminators to fire. ``Discr1" is the discriminator active inside the TPCW (outside the SW) for tagging cosmics. ``Discr3" is the discriminator active inside the BGW for beam gate events. When certain discriminator firing conditions are met (implemented in the FPGA on the PMT FEM), samples of various sizes (20 for tagging cosmics, 1500 for the SW) are read out from a delay memory on the FPGA to a DRAM memory portion of the FEM. These samples are read out through a separate ``trigger" data stream to DAQ machines if a trigger is issued from the trigger board. These samples are also read out through a continuous ``supernova" data stream to DAQ machines where they are stored for a few hours and then discarded pending a supernova trigger alert from the SNEWS system ~\cite{SNEWSref}.

\section{PMT Trigger Design and Implementation}
PMT triggers are formed in the FPGA on the PMT FEM. They are formed by implementing summed pulse amplitudes over multiple PMTs, coincidence conditions on multiple PMTs, and possibly delayed coincidences for a Michel-type trigger. All triggering is done on the ``high gain" PMT channel currently, though triggering also on the ``low gain" channel is possible. All PMT triggers formed are sent to the trigger board which then initiates the readout of all the TPC and PMT crates through the ``trigger" data stream. 
\subsection{Other Triggers}
There are other types of triggers available as input to the trigger board. All triggers input to the trigger board are "OR'd" logically. Other trigger types include but are not limited to a cosmic PMT trigger active in the OUTW to generate a clean sample of cosmics, DAQ-issued calibration triggers, and a random trigger.

\acknowledgments
The author thanks the organizers of LIDINE2013.

\end{document}